# Continuous-streaming high-speed two-photon dual-comb LiDAR with free-running lasers

BENJAMIN H. FORMAN,[1,*] HOLLIE WRIGHT,[1] HANNA OSTAPENKO,[1] AND DERRYCK T. REID[1]

[1] *Institute of Photonics and Quantum Sciences, School of Engineering and Physical Sciences, Heriot-Watt University, Edinburgh EH14 4AS, UK*
*\*bf35@hw.ac.uk*

**Abstract:** Dual-comb distance metrology combines sub-µm precision with absolute distance measurement over meter-scale non-ambiguity ranges. The most common interferometric implementation requires sophisticated approaches to achieve continuous streaming due to the need for digitization at near-gigasample/s rates. Here, we exploit the naturally low data burden of two-photon dual-comb LiDAR to demonstrate continuously-streamed distance metrology at 11.5 kHz, reaching a precision of nearly 1 µm in an averaging time of 10 ms. Using free-running 500 MHz Er,Yb:glass femtosecond lasers we achieve sampling rates >10 kHz, sufficient to demonstrate the capture of a four-minute audio track from the displacement of a loudspeaker-mounted mirror. The ability to perform high-speed yet low-data-burden distance metrology using unstabilized lasers presents exciting opportunities for diverse applications in industrial machine-tool control and calibration.

## 1. Introduction

As dual-comb techniques have evolved, systems capable of continuously streaming spectral or spatial measurements are becoming of increasing interest because they enable practical real-time acquisition by managing the conversion of dual-comb cross-correlations to data on the fly. Approaches for high-speed dual-comb spectroscopy have employed field-programmable gate array (FPGA) implementations [1] to enable real-time coherent-averaging procedures [2,3], and similar methods have been applied to dual-comb ranging for peak-detection of the interferograms [4] and to extend the non-ambiguity range [5]. More recently, long-range dual-comb distance metrology with real-time signal processing and target tracking has been reported using a graphics processing unit (GPU)-accelerated algorithm [6]. In the context of these computationally sophisticated solutions, two-photon dual-comb LiDAR [7,8] offers advantages over classical interferometric dual-comb measurements [9], including: simple envelope-like signals that are insensitive to the stability of the laser carrier-envelope offset frequency; self-calibrating measurements when $f_{\text{rep}}$ is known; wavelength and polarization insensitivity; and a low data burden from replacing analog digitization with time-stamping of two-photon cross-correlations. Similar benefits have been obtained in a dual-comb ranging scheme by using Type-0 sum-frequency mixing in a PPLN crystal [10] and Type-2 second-harmonic generation in a PPKTP crystal alongside FPGA-based acquisition [11].

Scaling dual-comb measurements to higher acquisitions rates requires increasing the difference in the comb repetition frequencies, $\Delta f_{\text{rep}}$, however the maximum rate is limited in the interferometric implementation by radio-frequency aliasing [9] and in two-photon dual-comb LiDAR by related temporal-overlap considerations [7]. In each case, the maximum $\Delta f_{\text{rep}}$ scales as $f_{\text{rep}}^2$, making high-repetition-rate sources preferable for high-speed sampling. Here, by employing two free-running Er,Yb:glass oscillators, each with $f_{\text{rep}} \approx 540$ MHz, we demonstrate sustained continuous streaming of absolute distance measurements at up to 11.5 kHz, increasing to 20 kHz for shorter durations, limited only by the on-board memory of the microcontroller we used.

## 2. Er,Yb:glass laser sources

The two nearly identical Er,Yb:glass lasers were configured in the 540 MHz cavity geometry shown in Fig. 1(a), with single-mode pump light from a 500 mW fiber-pigtailed 980 nm laser diode entering through one of the curved mirrors as shown. The gain medium was a 2 mm thick Brewster-angled plate of Er,Yb:glass, with Er and Yb dopings of 0.8% and 21% respectively. Both lasers were pumped at 980 nm by a single-mode fiber-pigtailed laser diode (3SP Technologies 1999CMB), which was coupled into the cavity by a combination of a 4.2 mm focal length aspheric collimator and a 35 mm focal length plano-convex spherical lens. The cavity mirrors were coated for high reflectivity (R > 99.8%) from 1500 nm–1600 nm and had high transmission at the pump wavelength.

Shown in Fig. 1(b), the lasers operated at different wavelengths, with one using a commercial SESAM (BATOP SAM-1550-3-4ps) with a loss profile that enabled modelocked operation at 1530 nm, while the other used a custom low-loss SESAM designed for 1557 nm operation [12] As introduced in [13], implementing two-photon dual-comb LiDAR using wavelength-separated probe and local-oscillator (LO) pulses avoids interference artifacts, improving cross-correlation quality. The use of different SESAMs also resulted in different modelocked output powers, with 13 mW produced at 1530 nm and 60 mW at 1557 nm.

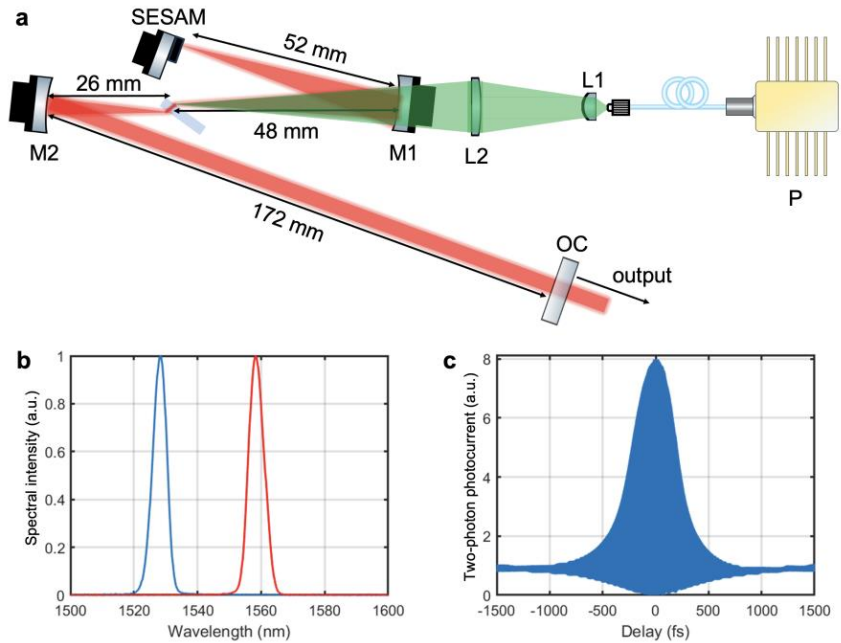

Fig. 1. (a) Cavity geometry of the 540 MHz Er,Yb:glass laser, with the 980 nm pump beam shown in green and the resonant 1530 nm or 1557 nm light shown in red. OC: output coupler; SESAM: semiconductor saturable absorber mirror; L1: aspheric collimator, focal length 4.2 mm; L2: focusing lens, focal length 35 mm; M1: plano-concave mirror, radius 50 mm; M2, plano-concave mirror, focal length 50 mm; P: 980 nm pump laser. (b) Spectra of both lasers each with a FWHM bandwidth of approximately 4.9 nm . (c) Two-photon autocorrelation of the 1557 nm laser.

A representative autocorrelation from the 1557 nm laser is shown in Fig. 1(c), indicating pulse durations of 534 fs. The duration-bandwidth product of 0.32 assuming $sech^2(t)$ pulses indicates that the pulses are transform-limited. The 1530 nm laser produced pulses with comparable durations.

## 3. Two-photon dual-comb LiDAR design

The two-photon dual-comb LiDAR scheme is shown in Fig. 2. Pulses from the more powerful 1557 nm laser were used as the probe, and were divided between the target and reference reflectors before being polarization multiplexed with local-oscillator (LO) pulses from the 1530 nm laser. The two-photon detector (Fig. 2, TPA) was a fiber-pigtailed laser diode (Luminent, MRLDFC010), which provided considerably higher two-photon absorption responsivity than a simple photodiode [14,15] along with convenient connectorization. The dual-comb signal obtained directly from the detector was a sequence of 2 mV cross-correlations, whose amplitudes were increased to 150 mV using a high-speed current amplifier (Femto HCA-20M-100K-C), whose 20 MHz bandwidth also suppressed the pulse repetition frequency, $f_{\text{rep}}$. A constant fraction discriminator (CFD) then transformed these cross-correlations into CMOS-compatible 3.3 V amplitude pulses, with rising edges whose positions were insensitive to the pulse amplitude.

We configured the system for dual-comb ranging at sampling rates from 10 kHz – 20 kHz, determined by the difference in the pulse repetition frequencies of the lasers, $\Delta f_{\text{rep}}$. Figure 2(b) shows a representative radio-frequency spectrum of the combined lasers (resolution bandwidth 10 Hz), operating with $f_{\text{rep}}$= 540 MHz and $\Delta f_{\text{rep}}$= 11.73 kHz.

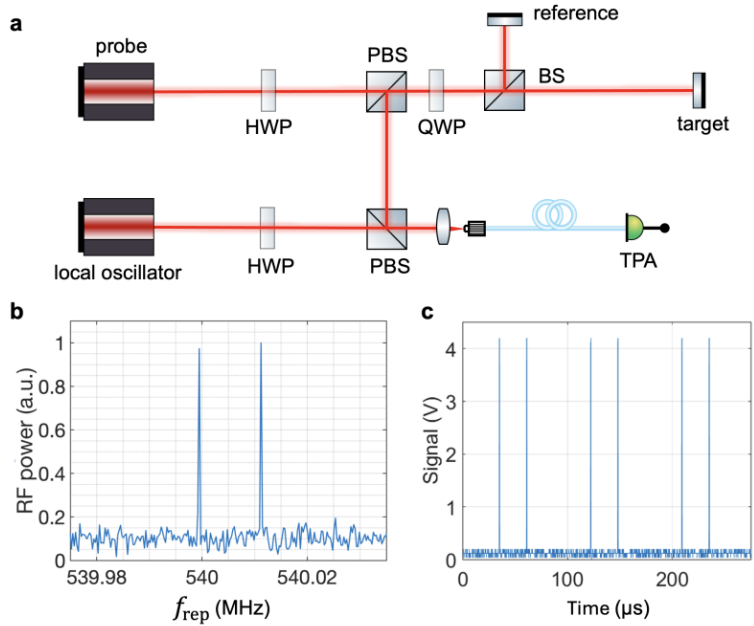

Fig. 2. (a) Two-photon dual-comb LiDAR scheme, showing polarization multiplexing of the probe and local oscillator pulses. BS: non-polarizing beam splitter; PBS: polarizing beamsplitter; HWP: half-wave plate; QWP: quarter-wave plate; TPA: two-photon absorption detector. (b) Representative a radio-frequency spectrum of the combined lasers. (c) Example of the dual-comb cross-correlations after signal conditioning (see text).

In previous work, we used a 600 MHz Teensy microcontroller to directly time-stamp each of the two-photon cross-correlations, leading to a dataset containing the time intervals between successive cross-correlations [7]. While this approach is simple and robust, it is limited by the clock speed of the microcontroller to a precision of 1.67 ns, imposing a constraint on the distance metrology accuracy. Here, we introduce an alternative approach based on the commonly available TDC7200 time-to-digital converter (TDC) [16], which has a timing resolution of 55 ps. This TDC is designed for time-of-flight applications in LiDAR and performs the function of a stopwatch by measuring the elapsed time between a START pulse and a STOP pulse.

The timing measurement scheme employed two TDC7200 devices, both synchronized to a common 16 MHz clock, and is shown in Fig. 3. A D-type flip-flop was used to demultiplex the LO-reference (R) and LO-target (T) cross-correlations (Fig. 2(a)) into two separate measurement channels. Isolating the cross-correlations in this way allowed us to use two TDCs to, respectively, measure the short reference-target gap and the longer target-reference gap illustrated schematically in Fig. 3.. The timing data from the TDCs were streamed continuously over an SPI interface to a Teensy 4.0 microcontroller. This low data burden allowed USB acquisition to a local computer at up to 12 kSa/s with a data burden of 64 bits per sample (2 unsigned integers). The asymmetry in the latencies of the two channels was easily removed by a simple calibration step that supplied the input to the timing system with a sequence of 50 ns pulses from a clock source. While not demonstrated here, this system could easily be extended to an atomically-referenced measurement by using an external GPS-referenced clock signal for the TDCs.

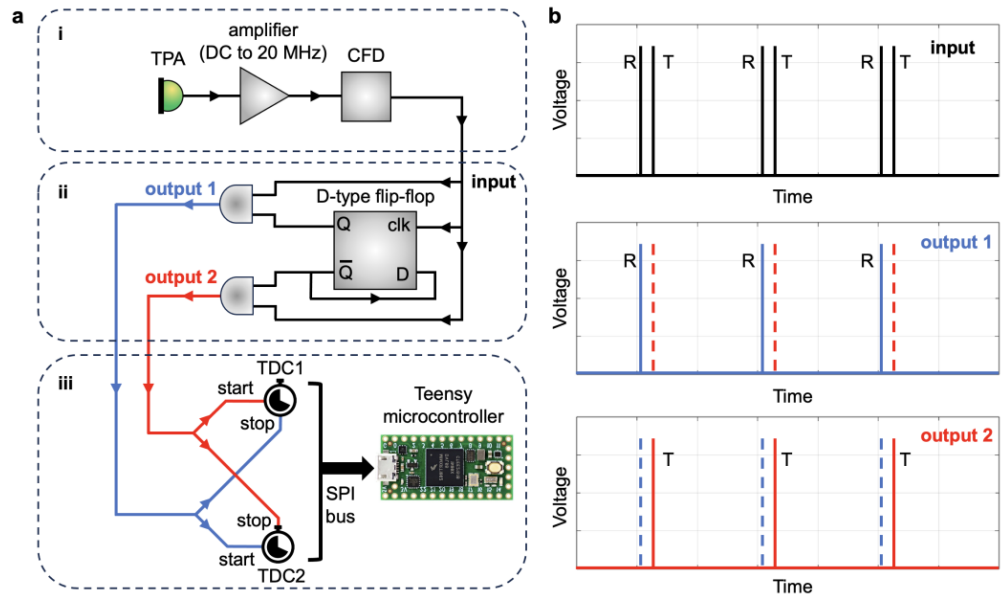

Fig. 3. Timing measurement scheme. (a) Signal amplification and conditioning (see text) transforms the two-photon cross-correlations into CMOS-compatible pulses (i), which are divided into two channels. One channel is used to toggle a flip-flop, the outputs of which alternately enable and disable two AND gates to create a demultiplexer (ii). The other channel supplies the second input to the AND gates. The outputs of the AND gates (iii) are the demultiplexed LO-reference (R) and LO-target (T) cross-correlations, which are used respectively as start / stop triggers for TDC1 and stop / start triggers for TDC2. (b) Illustrative inputs and outputs of the demultiplexing scheme.

## 4. Metrology characterization

### 4.1 High speed acquisition to microcontroller memory

The maximum data acquisition rate achieved without missing samples was 20 kHz, however this could not be done continuously because the additional overhead of data transfer over the USB interface would not support measurements with a delay between the target and reference pulses of < 10 μs. Instead, the measurement involved saving the data to an array within the microcontroller RAM, simplifying acquisition but at the expense of limiting the maximum number of samples which could be acquired. All measurements were recorded with the lasers free-running, using no active control over $f_{\text{rep}}$, and only manually adjusting the cavity length of one laser to bring $\Delta f_{\text{rep}}$ into the correct range. Figures 4(a) and 4(b) show the directly measured reference-target and target-reference timing intervals (left), along with a histogram of the data (right). In Fig. 4(c) we show the sum of the measurements in Figs. 4(a) and 4(b), corresponding to the sampling period of the distance measurements, $1/\Delta f_{\text{rep}}$. The variations from the mean in the measurements in Figs. 4(a)–4(c) reflect the small drift in the repetition-rate differences of the free-running lasers.

The target-reference distance measurement, $d_i$, is obtained from a simple ratio [17,7], where $i$ is an index identifying a particular measurement, $v_g$ is the group velocity of light (assumed to equal $c$ for the purposes of our study) and $t$ is the timing interval reported between consecutive target-reference ($TR$) or reference-target ($RT$) cross-correlations:

$$d_i = \frac{v_g}{2f_{\text{rep}}} \cdot \frac{t_i^{TR}}{t_i^{TR} + t_i^{RT}}$$

Because $\Delta f_{\text{rep}}$ fluctuations scale $t_i^{TR}$ and $t_i^{RT}$ equally, the distance measurement benefits from strong common-mode rejection of repetition-rate variations. This configuration allowed for extremely rapid measurements which average down to a precision of 1 µm in 30 ms.

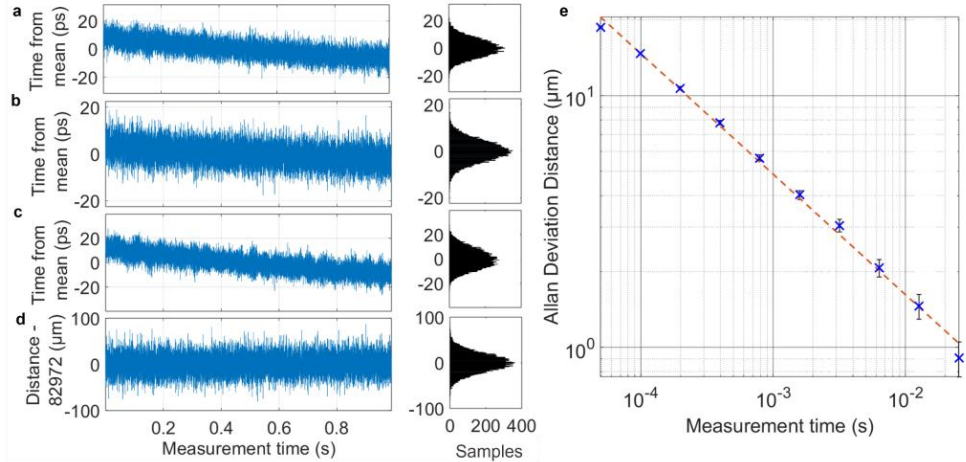

Fig. 4. (a, b) Reference-target and target-reference cross-correlation timing measurements running at 20 kHz for 1 second. (c) Timing period, indicating $\Delta f_{\text{rep}}$ fluctuations. (d) Distance between the target and reference reflectors, inferred from data in (a) and (b). (e) Allan deviation of the distance data in (d), showing averaging to 1 µm precision in 30 ms of observation time. The red line is a fit to the data showing the expected $1/\sqrt{\tau}$ behavior.

### *4.2 Continuous streaming to host PC*

To subvert the RAM limitation discussed in Section 4.1, a variation of the system was designed to continuously write measurements to the serial port of the Teensy microcontroller. The metrology performance of this system was evaluated for 17 seconds of continuously streaming data by recording a two-photon dual-comb LiDAR measurement with a target-reference path difference of 6.2 cm (single-pass) and a sampling rate of $\Delta f_{\text{rep}}$ = 11.5 kHz.

Figure 5, which is organized identically to Fig. 4, shows the LiDAR timing measurements over this much longer acquisition time, illustrating clearly their sensitivity to variations in $\Delta f_{\text{rep}}$ but also the common-mode rejection of these in the distance calculation, which exhibits Gaussian distributed noise with a standard deviation of 11.7 µm. Figure 5(e) shows the accompanying Allan deviation for the distance measurements in Fig. 5(d), indicating that a distance precision of 1 µm is reached after only 10 ms of averaging. This is faster than the 20 kHz measurement due to the slower cross correlation speed providing a lower timing uncertainty and as a result a lower standard deviation in the distance measurement.

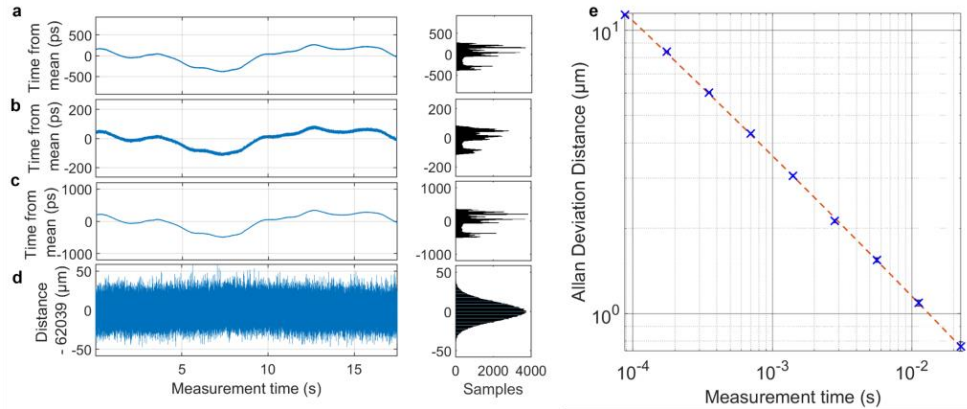

Fig. 5. (a)-(d) definitions as in Fig. 4, but with a measurement rate of 11.5 kHz and a data acquisition time of 17 seconds. (e) Allan deviation of the data in (d), showing averaging to a precision of 1 µm in a time of ~10 ms.

## 5. Continuous streaming of a dynamic target

To demonstrate the ability to apply two-photon dual-comb LiDAR to high-sampling-rate, low-data-burden dynamic measurements, we configured the Teensy microcontroller to continuously stream over the USB serial port to a host computer, enabling sustained sampling at $\Delta f_{\rm rep} \approx 10$ kHz. This configuration was used to digitize an audio track lasting for four minutes, by replacing the static target mirror in Fig. 2(a) with one mounted on a loudspeaker. The audio track, a low-pass-filtered version of Hoziers "Too Sweet", was encoded in the displacement of the mirror, which was measured at $\Delta f_{\rm rep}$ = 9.3 kHz using the dual-comb ranging system.

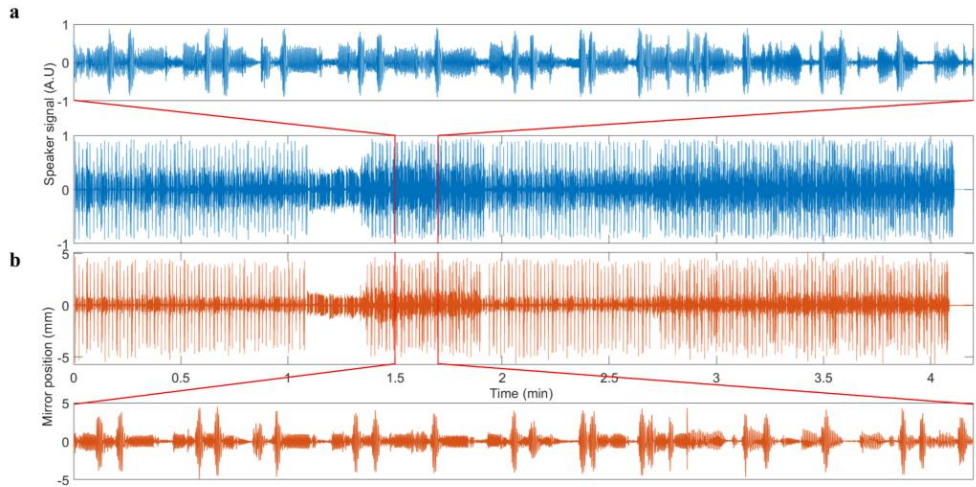

Fig. 6. (a) Original (blue) including a 6 dB per octave low pass filter to account for the high bass response of the speaker, and (b) dual-comb-sampled (red) audio waveforms, each with a 12-second zoom. The measurements were made with $\Delta f_{\rm rep}$ = 9.3 kHz.

The original waveform is shown in Fig. 6(a), while Fig. 6(b) shows the dual-comb measurement. In some high-frequency sections of the audio track, the dual-comb data exhibit a lower amplitude than the original waveform, reflecting the reduced high-frequency response of the loudspeaker after loading with a mirror. These data directly illustrate the capability of the system for sustained continuous streaming, with no fundamental limit on the size of the recorded dataset except for the storage capacity of the local computer.

## 6. Conclusions

These results demonstrate the potential for high-speed dual-comb LiDAR made possible by the high repetition rates available from dual Er,Yb:glass lasers, which can also be implemented in a single cavity configuration [18] permitting a simplified dual-comb ranging scheme [19]. Our system has a fundamental optical limit of $\Delta f_{\text{rep}} \leq 133$ kHz, which is not reached because of two bottlenecks. The first is the SPI bus from the TDCs and the Teensy, which supports a data transfer rate of 20 Mbps, however each distance sample requires at least 640 bits of data resulting in a hard limit of 31.25 kS/s. The continuous streaming speed is limited by the transfer time of the Teensy, which in turn is determined by the speed of the USB serial bus, but even for the relatively slow 480 Mb/s rates of USB 2.0, this is not expected to be a major constraint. Even in its present form, the continuous streaming capability is sufficient to be applied in the real-time calibration and control of robotic and machine-tool systems where inferring the position of a component from encoders alone may introduce considerable uncertainty.

**Funding.**

Engineering and Physical Sciences Research Council (EP/Y011422/1). Royal Academy of Engineering (RCSRF2223-1678). Renishaw plc.

**Disclosures.**

The authors declare no conflicts of interest.

**Data Availability.**

Data underlying the results presented in this paper are not publicly available at this time but may be obtained from the authors upon reasonable request.